\documentclass{mem}
\usepackage{amsmath} 
\usepackage{natbib}\usepackage{txfonts}\usepackage{balance}
\usepackage{flushend}
\usepackage{subcaption}
\usepackage{graphicx}
\usepackage[breaklinks,pdftex]{hyperref}
\usepackage[utf8]{inputenc} 
\usepackage[T1]{fontenc}

\begin{document}

\title{Search for Short Timescale Variability in PG1553+113 with LST-1 of CTAO}


\author{
A. Ruina\inst{1} \and 
H. Luciani\inst{2,3} \and 
E. Prandini\inst{4} \and
the LST Collaboration of the CTAO}

\institute{
Istituto Nazionale di Fisica Nucleare, Sezione di Padova, Italy \\\email{arshia.ruina@pd.infn.it}
\and
Dipartimento di Fisica, Università di Trieste, Italy
\and
Istituto Nazionale di Fisica Nucleare, Sezione di Padova, Italy
\and
Dipartimento di Fisica e Astronomia "Galileo Galilei" (DFA), Università degli Studi di Padova, Italy
}

\authorrunning{Ruina et al. (LST Collaboration)}

\titlerunning{Short Timescale Variability in PG1553+113}

\date{Submitted to the proceedings of the 8th Heidelberg International Symposium on High-Energy Gamma-Ray Astronomy ($\gamma$-2024)}

\abstract{
PG 1553+113 is a high-frequency peaked BL Lac object (HBL), with redshift 0.433, detected with the current generation of IACTs (Imaging Atmospheric Cherenkov Telescopes) up to $\sim$ 1 TeV. Interestingly, the continuous $\gamma$-ray lightcurve collected by Fermi-LAT since 2008 showed a signature of a periodic modulation of $2.18 \pm 0.08$ years at energies above 100 MeV and 1 GeV. In addition, the source shows clear variabiliy down to day-scale in all bands. XMM-Newton data recently showed rapid variability in the X-ray band down to $2.4 \pm 0.7$ ks.

Short-timescale (sub-hour) variabilities are a key observable to constrain the size of the photon-emitting region inside the blazar jet. The LST-1 (first protoype of the Large-Sized Telescope) of the CTAO (Cherenkov Telescope Array Observatory) is located on Roque de los Muchachos in La Palma, Spain. With its high sensitivity at low energies (20-150 GeV), it provides a unique opportunity to investigate such phenomena. In 2023, the source had a very bright flare that triggered LST-1 and multi-wavelength data campaigns. In this study, we present the results of this observation campaign, in particular, the search for short-timescale variabilities.}

\maketitle{}

\section{Introduction}

Blazars are a class of active galactic nuclei (AGNs) that have their relativistic jets almost aligned with our line of sight on the Earth \citep{Urry_1995}. The non-thermal emission is strongly beamed and spreads across the entire electromagnetic spectrum, from radio waves to very high energy (VHE) $\gamma$-rays, often showing variability in different timescales such as intra-day variability, or IDV (minutes to hours), short-term varibility (several days to a few months) and long-term variability (several months to a few years). The underlying mechanisms for variability are a subject of ongoing research e.g. physical processes in the jet or in the accretion mechanism (\citep{Marscher_2016, Raiteri_2017} and references therein).

Their spectral energy distributions (SEDs) are characterised by two peaks, one in low energies, typically between the far-infrared and the soft X-ray bands and another in high energies, between the hard X-ray and the $\gamma$-ray bands. The low-energy peak is largely explained by synchrotron emission from relativistic electrons moving in a magnetic field within the jet. The origin of the high-energy peak is described by different mechanisms in leptonic and hadronic models. In leptonic models, it is attributed to inverse-Compton scattering of the synchrotron photons off the relativistic electrons (synchrotron self-Compton, or SSC model) or of the thermal photons from surrounding structures like the broad-line region or dusty torus (external Compton, or EC model) (references within \citep{Liu_2023}). In hadronic models, it is attributed to proton synchrotron radiation \citep{Aharonian_2000} or photohadronic interactions \citep{Petropoulou_2016}. 

In this work, the high-frequency peaked BL Lac object (HBL; a blazar with a featureless continuum in its optical spectrum and whose synchroton peak lies at $>10^{15}$ Hz) called PG1553+113, discovered by the Palomar-Green survey of UV-excess stellar objects \citep{Green_1986}, is under study. It lies at a redshift of $z=0.433$ \citep{Jones_2022} with (ICRS) coordinates $238.93^{\circ}$, $11.2^{\circ}$ (R.A.,Dec.). So far, VHE $\gamma$-ray emission was discovered independently and almost simultaneously by H.E.S.S. \citep{Aharonian_2006} and MAGIC \citep{Albert_2007} in 2005 and up to 1 TeV with current generation of imaging air-shower Cherenkov telescopes (IACTs) \citep{Aharonian_2008}. The flux in GeV energies is variable, as is typical of TeV blazars \citep{Aleksic_2015}. Also interesting to note, in 2015, Fermi-LAT reported a quasi-periodicity in its variability, with a period of $2.18\pm0.08$ years for energies $E>100$ MeV and $E>1$ GeV \citep{Ackermann_2015}. 

More recently, XMM-Newton detected the presence of IDV on the timescale of $2.4\pm0.7$ ks \citep{Dhiman_2021} for energies 0.3-10 keV. Such variabilities have not yet been detected at TeV energies for this source, which motivates further studies, as it could be an intrinsic property of the source or due to observation periods that are too short (1-2 hours). Moreover, sub-hour variablities are a key observable to probe the small spatial structures of the jet e.g. constraining the size of the photon-emitting region in the relativistic jet \citep{Marscher_2010} and acceleration mechanisms in the jet e.g. in shocks or through magnetic reconnection \citep{Giannios_2010}.

\section{The LST-1 Instrument}

The Cherenkov Telescope Array Observatory (CTAO) will be a next-generation ground-based observatory for $\gamma$-rays in VHEs. Three classes of telescopes will be part of the array to cover a broad range of energies from 20 GeV to 300 TeV with an order of magnitude improvement in sensitivity. LST-1 is the protoype of the large-sized telescope, located in the northern site of CTAO, on Roque de los Muchachos Observatory in La Palma, Spain. It was inaugurated in 2018 and has been taking data since November 2019 \citep{Abe_2023}. With a mirror dish diameter of 23 m, LSTs have an optimised sensitivity for CTAO’s low energy range i.e. 20-150 GeV, which gives us a unique opportunity to investigate blazar variabilities.

\begin{figure*}
\resizebox{\hsize}{!}{\includegraphics[clip=true]{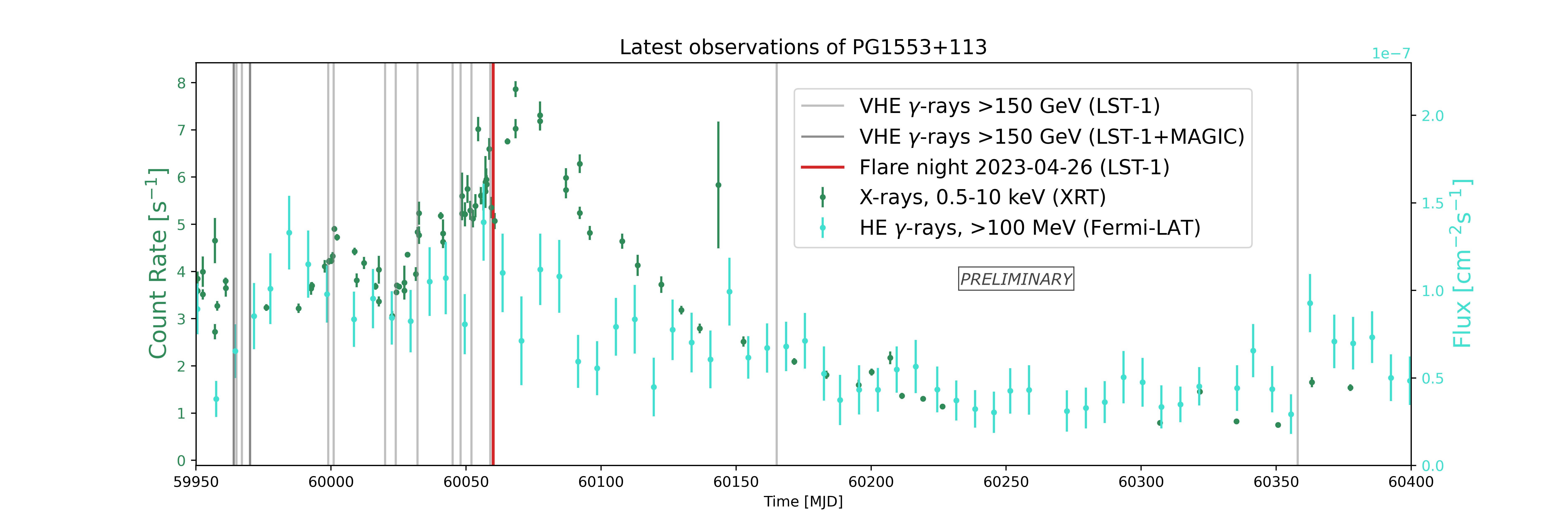}}
\caption{\footnotesize{X-ray count rates from Swift-XRT, HE $gamma$-ray flux from Fermi-LAT and dates of VHE $gamma$-ray observations from LST-1 and LST-1+MAGIC, triggering on the high states. LST-1's observation of the flare on April 26th, 2023, indicated by the red line, was triggered on the MAGIC monitoring data during the peak of the Fermi-LAT periodicity.}}
\label{fig:latest_obs}
\end{figure*}

PG 1553+113 has been detected by LST-1 with a high significance since 2021 (publication in preparation). The source is highly variable, with an average emission state at around 10-20\% Crab units. In 2023, the source was in a high state (coinciding with the peak of the $\sim$2-year Fermi-LAT periodicity) and LST-1 made a long-exposure observation of a flare on April 26th (see figure \ref{fig:latest_obs}).

\section{Observations and Analysis}

The signal extraction, calibration and reconstruction of events for the data collected by LST-1 during the flaring state of P1553+113 that occured on April 26th, 2023 was done using the official software \texttt{cta-lstchain}\footnote{\url{https://doi.org/10.5281/zenodo.10849683}}. Monte-Carlo (MC) simulations and the derived instrument response functions (IRFs) were obtained using the procedure outlined in \citep{Abe_2023}.
A source-independent approach is used in this analysis, meaning that no assumptions are made about the location of the source. Data were taken in 20-minute runs in the ``wobble mode'', where the source is located at an offset of $0.4^{\circ}$ from the camera center. Standard data-cleaning procedures were implemented, and additional requirements of each observation run to be longer than 30 s and at zenith angles $<52^{\circ}$ were placed. This resulted in 15 runs with a livetime of 3.67 h. Hereafter, the high-level data analysis is carried out using the \textit{GammaPy} package \citep{gammapy:2023}.

\begin{figure}
\resizebox{\hsize}{!}{\includegraphics[clip=true]{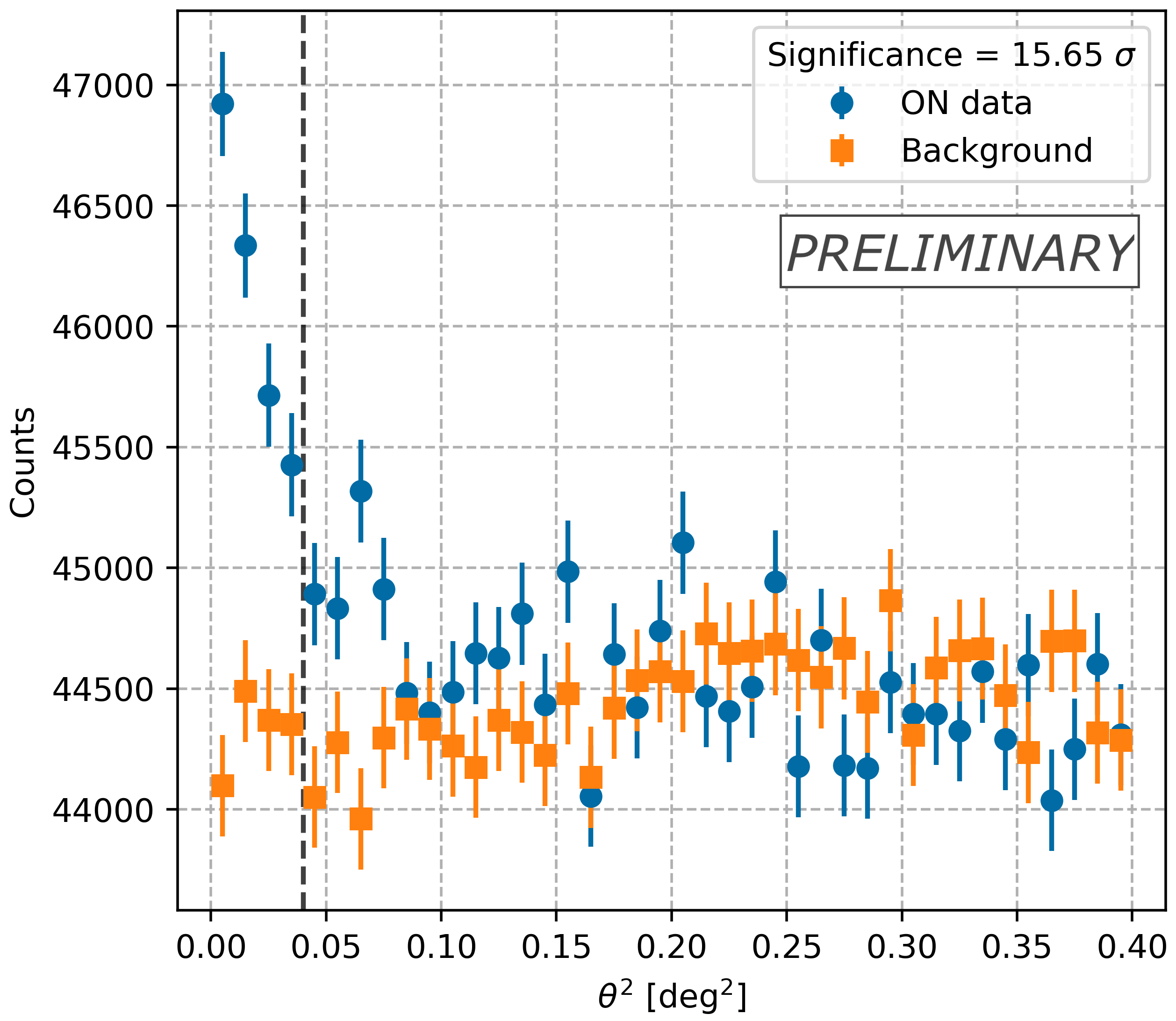}}
\caption{\footnotesize{Distribution of the squared angular distance, $\theta^{2}$, of the $\gamma$-ray excess from the nominal source position for on-source events and normalized off-source (background) events. The dashed vertical line at $\theta^{2}=0.04^{\circ 2}$ represents the on-source integration region. (Reconstructed energy is between 10 GeV and 10 TeV.)}}
\label{fig:thetasq}
\end{figure}

\begin{figure}
\resizebox{\hsize}{!}{\includegraphics[clip=true]{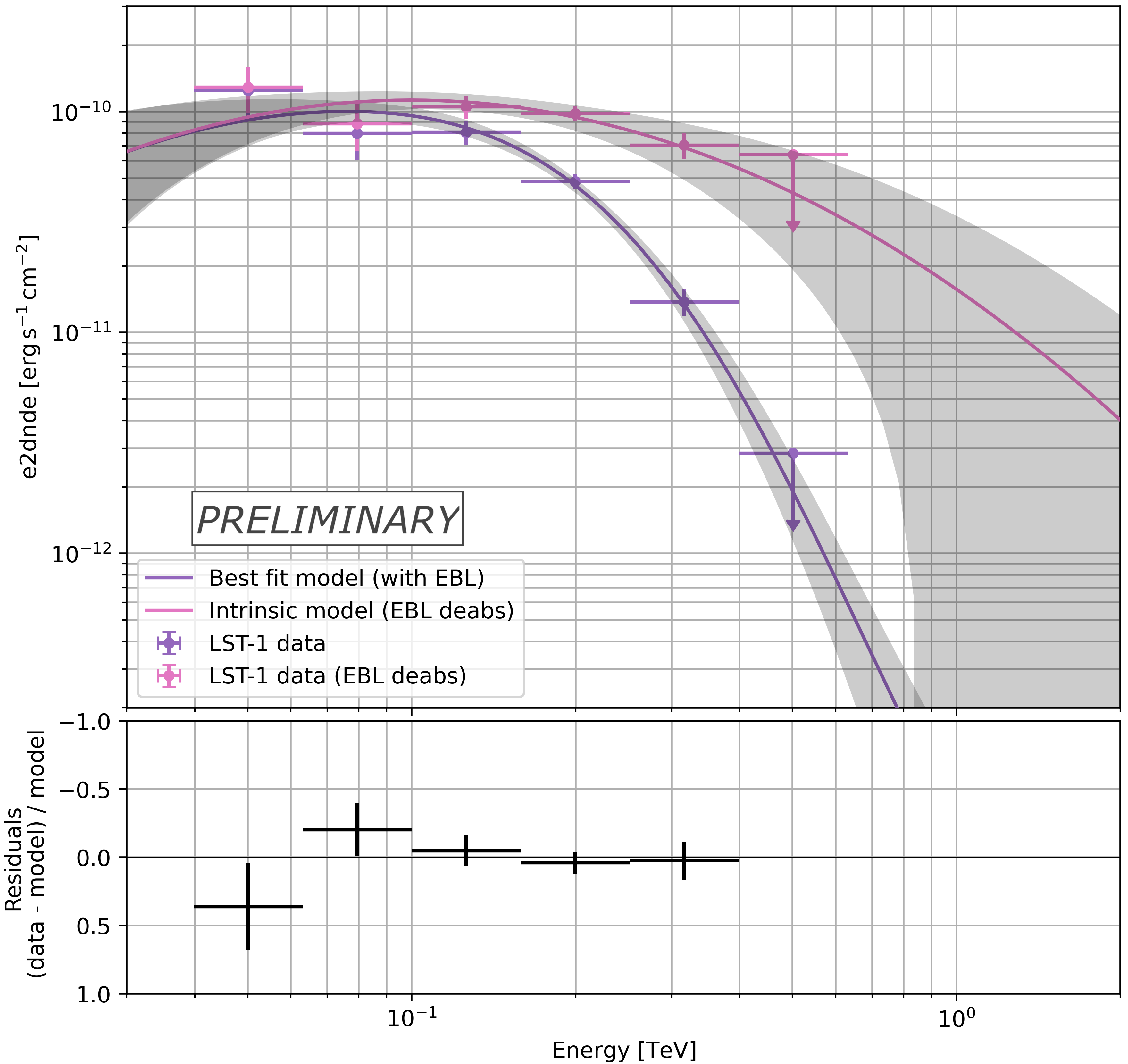}}
\caption{\footnotesize{The spectral energy distribution is constructed by fitting the data to a log-parabola function, where the intrinsic spectrum is obtained after correcting for EBL absorption \citep{Dominguez_2011}. The detected energy ranges from 40 GeV to 400 GeV.}}
\label{fig:sed}
\end{figure}

With $\theta$ defined as the angular distance between the reconstructed direction and the source position, the $\theta^{2}$-distribution is shown in figure \ref{fig:thetasq}. It is consistent with the hypothesis of detection of a point source with a significance $>15\sigma$. The SED is shown in figure \ref{fig:sed} for the observed data as well as for the extra-Galactic background light (EBL) de-absorbed scenario (i.e. intrinsic spectrum). An analytical function was used to mathematically model the SED:

\begin{equation}
	\Phi(E) =  \Phi_{0} \bigg( \frac{E}{E_{0}} \bigg) ^ {-\alpha -\beta \log \big(\frac{E}{E_{0}} \big)}
\end{equation}

This is the so-called log-parabolic function that is commonly used to fit the spectral distributions of non-thermal sources like blazars. The best fit parameters obtained for the observed spectral fit are as follows:
\begin{itemize}
\item reference energy, $E_{0} = 120$ GeV, 
\item amplitude, $\Phi_{0} = (4.8 \pm 0.4) \times 10^{9}$ TeV\textsuperscript{-1} s\textsuperscript{-1} cm\textsuperscript{-2},
\item alpha, $\alpha = 2.14 \pm 0.15$, and,
\item beta, $\beta = 0.37 \pm 0.24$.
\end{itemize}
%
%

\begin{figure}
\resizebox{\hsize}{!}{\includegraphics[clip=true]{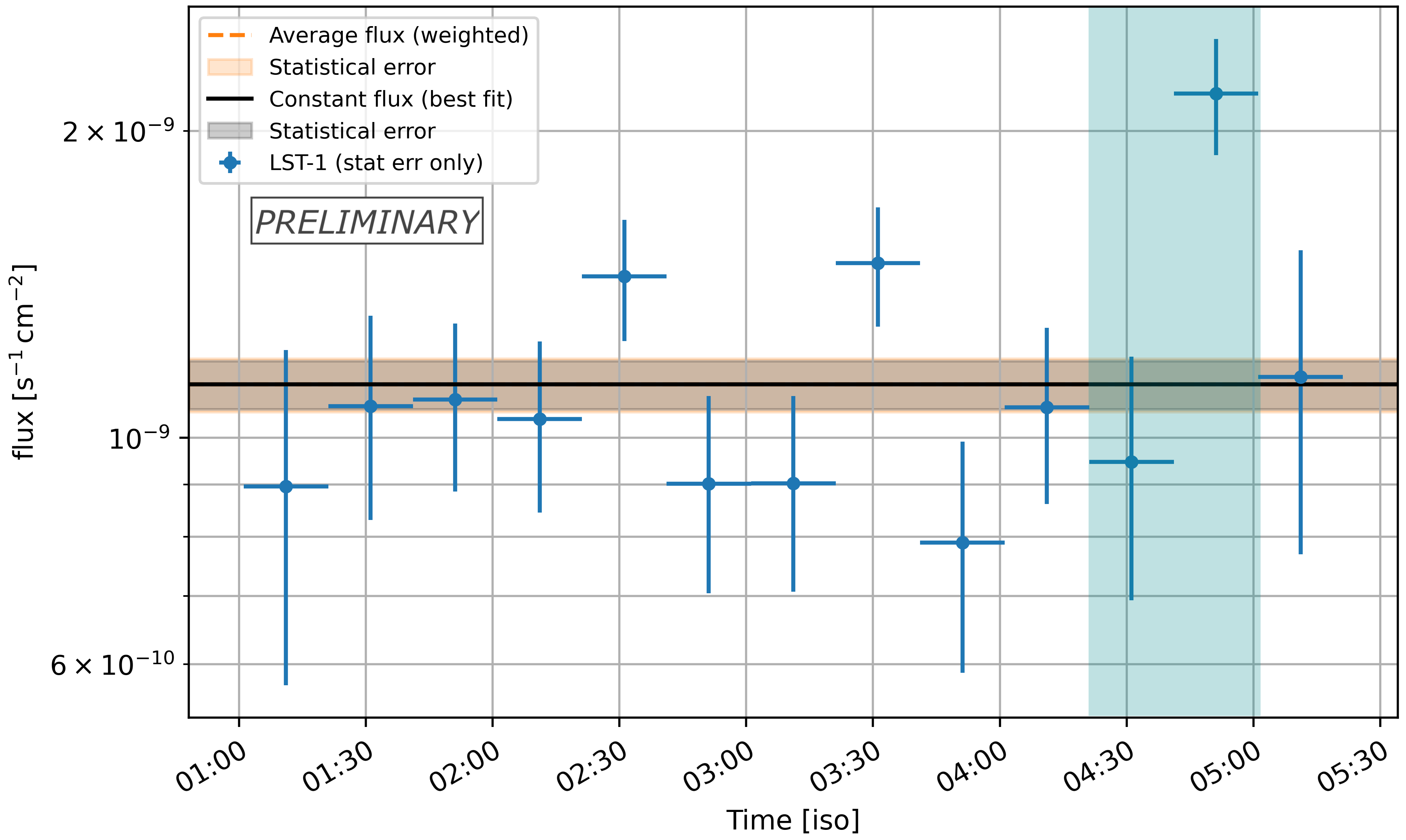}}
\caption{\footnotesize{The light curve is constructed for the flaring state of PG1553+113 as observed by LST-1 on April 26th, 2023. The weighted average of the flux and the best fit to a constant value of $(1.12 \pm 0.06) \times 10^{9}$ s\textsuperscript{-1} cm\textsuperscript{-2}, with $\chi^{2}=26.2$, are also shown. Only statistical errors have reported here. The teal band indicates the point in time at which the variability timescale is computed (see text).}}
\label{fig:lc}
\end{figure}

These flux points are used to create the light curve (LC), in bins of 20 minutes, with an energy threshold set at 50 GeV (3 times lower than the threshold for the MAGIC telescopes due to the higher sensitivity of LSTs), and is shown in figure \ref{fig:lc}. A preliminary analysis to search for any hint of variability in the LC was conducted and is presented in the following.

\subsection{Characterisation of the Variability}

The uncertainties in a LC that arise from measurement errors contribute to an additional variance. As such, the sample variance is, essentially, the sum of the intrinsic variance from the source and this ``excess'' variance. Subtracting the latter from the sample variance and normalising it gives us an estimator of the intrinsic source variance. This is defined as \citep{Edelson_2002,Vaughan_2003}

\begin{equation}
	F_{\text{var}} = \sqrt{\frac{S^{2}- \langle \sigma^{2}_{\text{err}} \rangle } {{\langle x \rangle}^{2} } }
\end{equation}

where, $S^{2}$ is the sample variance, $\langle \sigma^{2}_{\text{err}} \rangle$  is the mean squared error, $\langle x \rangle$ is the mean of the measurements and $F_{\textrm{var}}$ is the fractional RMS variability amplitude, the statistic used to characterise the intrinsic variability in the source. From the LC presented in figure \ref{fig:lc}, a variability amplitude of $F_{\textrm{var}} (\%) = 24.0\pm6.9$ is obtained with a 3.5 $\sigma$ significance.

\subsection{Variability Timescales}
\label{subsec:variability_timescales}

The shortest variability timescale is computed using the formula for flux doubling/halving \citep{Foschini_2011}

\begin{equation}
	F(t_{1}) = F(t_{2}) 2^{(t_{1}-t_{2})/\tau} 
\end{equation}

where $\tau$ is the characteristic halving/doubling time-scale and $F(t_{1})$ and $F(t_{2})$ are the fluxes of the LC at times $t_{1}$ and $t_{2}$, respectively. The shortest timescale was found to be $0.99\pm0.17$ ks at/after 2023-04-26 04:31:09.184 UTC (as indicated by the teal band in figure \ref{fig:lc}).

\subsection{Constraining the Radius of the Photon-Emitting Region}

An upper limit on the radius of the photon-emitting region inside the relativisitic jet of the source is calculated using

\begin{equation}
	R \leq \frac{c t_{\text{var}} \delta}{1+z}
\end{equation}

where $c$ is the speed of light in vacuum, $t_{\text{var}}$ is the variability timescale computed in section \ref{subsec:variability_timescales}, $\delta$ is the Doppler factor, taken to range between 11 and 35 \citep{Dhiman_2021} and $z$ is the redshift. The upper limit on the radius is within the range of $(0.23-0.73)\times10^{15}$ cm.

\section{Discussion}

A preliminary analysis of the data collected by the LST-1 telescope during the 2023 high state of the BL Lac object PG1553+113 is presented. It is a demonstration of the powerful capability of the next-generation IACTs and what can potentially be achieved using the final array of 4 LSTs that will become part of the northern site of the CTAO. The LC from the source of interest was constructed with an energy threshold as low as 50 GeV, using the offical analysis pipeline and \textit{GammaPy}, and the machinery to search for variability in the flux was implemented. We report a variability amplitude of $F_{\textrm{var}} (\%) = 24.0\pm6.9$ with a 3.5 $\sigma$ significance, hinting at the presence of IDV with a timescale of $0.99\pm0.17$ ks, for a light curve spanning nearly 4.5 hours. We use this to provide an upper limit range on the radius of the photon-emitting region in the jet as $(0.23-0.73)\times10^{15}$ cm. The work, however, is still under review.

Exploiting the $\sim$2-year periodicity in high-energy $\gamma$-rays, as predicted by Fermi-LAT, we expect to observe the source again in its flaring state in 2025 and collect more data to better probe the presence of IDV. Discussions on future observations, including multi-wavelength campaigns, and also during the low-state of the source, are currently ongoing. It is highly anticipated that the higher sensitivity of the upcoming CTAO, especially in lower energies, will help us understand not only this source and its properties better but also blazars, in general, some of the most energetic objects in the Universe.

\begin{acknowledgements}
We gratefully acknowledge financial support from the agencies and organizations listed here: \url{https://www.ctao.org/ for-scientists/library/acknowledgments/}.
\end{acknowledgements}

\bibliography{bibliography}

\begin{thebibliography}{22}
\expandafter\ifx\csname natexlab\endcsname\relax\def\natexlab#1{#1}\fi

\bibitem[{Abe {et~al.}(2023)Abe, Abe, Abe, Aguasca-Cabot, Agudo, Crespo,
  Antonelli, Aramo, Arbet-Engels, Arcaro, Artero, Asano, Aubert, Baktash,
  Bamba, Larriva, Baroncelli, de~Almeida, Barrio, Batkovic, Baxter, González,
  Bernardini, Bernardos, Medrano, Berti, Bhattacharjee, Biederbeck, Bigongiari,
  Bissaldi, Blanch, Bonnoli, Bordas, Borghese, Bulgarelli, Burelli, Buscemi,
  Cardillo, Caroff, Carosi, Cassol, Cauz, Ceribella, Chai, Cheng, Chiavassa,
  Chikawa, Chytka, Cifuentes, Contreras, Cortina, Costantini, D’Amico,
  Dalchenko, Angelis, de~Bony~de Lavergne, Lotto, de~Menezes, Deleglise,
  Delgado, Mengual, della Volpe, Dellaiera, Depaoli, Piano, Pierro, Tria,
  Venere, Díaz, Dominik, Prester, Donini, Dorner, Doro, Elsässer, Emery,
  Escudero, Ramazani, Ferrara, Ferrarotto, Fiasson, Coromina, Fröse, Fukami,
  Fukazawa, Garcia, López, Gasbarra, Gasparrini, Geyer, Paiva, Giglietto,
  Giordano, Giro, Gliwny, Godinovic, Grau, Green, Green, Gunji, Hackfeld,
  Hadasch, Hahn, Hashiyama, Hassan, Hayashi, Heckmann, Heller, Llorente,
  Hirotani, Hoffmann, Horns, Houles, Hrabovsky, Hrupec, Hui, Hütten, Iarlori,
  Imazawa, Inada, Inome, Ioka, Iori, Ishio, Iwamura, Jacquemont, Martinez,
  Jurysek, Kagaya, Karas, Katagiri, Kataoka, Kerszberg, Kobayashi, Kong, Kubo,
  Kushida, Lainez, Lamanna, Lamastra, Flour, Linhoff, Longo, López-Coto,
  López-Moya, López-Oramas, Loporchio, Lorini, Luque-Escamilla, Majumdar,
  Makariev, Mandat, Manganaro, Manicò, Mannheim, Mariotti, Marquez, Marsella,
  Martí, Martinez, Martínez, Martínez, Marusevec, Mas-Aguilar, Maurin,
  Mazin, Guillen, Micanovic, Miceli, Miener, Miranda, Mirzoyan, Mizuno,
  Gonzalez, Molina, Montaruli, Monteiro, Moralejo, Morcuende, Morselli,
  Mrakovcic, Murase, Nagai, Nagataki, Nakamori, Nickel, Nievas, Nishijima,
  Noda, Nosek, Nozaki, Ohishi, Ohtani, Oka, Okazaki, Okumura, Orito,
  Otero-Santos, Palatiello, Paneque, Pantaleo, Paoletti, Paredes, Pech,
  Pecimotika, Peresano, Pérez, Pietropaolo, Pirola, Plard, Podobnik, Poireau,
  Polo, Pons, Prandini, Prast, Principe, Priyadarshi, Prouza, Rando, Rhode,
  Ribó, Rizi, Fernandez, Ruiz, Saito, Sakurai, Sanchez, Šarić, Sato,
  Saturni, Schleicher, Schmuckermaier, Schubert, Schussler, Schweizer, Arroyo,
  Silvia, Sitarek, Sliusar, Spolon, Strišković, Strzys, Suda, Sunada, Tajima,
  Takahashi, Takahashi, Takata, Takeishi, Tam, Tanaka, Tateishi, Tejedor,
  Temnikov, Terada, Terauchi, Terzic, Teshima, Tluczykont, Tokanai, Torres,
  Travnicek, Truzzi, Tutone, Uhlrich, Vacula, Vallania, van Scherpenberg,
  Acosta, Verguilov, Viale, Vigliano, Vigorito, Vitale, Voutsinas, Vovk,
  Vuillaume, Walter, Will, Yamamoto, Yamazaki, Yoshida, Yoshikoshi, Zywucka,
  Bernlöhr, Gueta, Kosack, Maier, \& Watson}]{Abe_2023}
Abe, H., Abe, K., Abe, S., {et~al.} 2023, The Astrophysical Journal, 956, 80

\bibitem[{Ackermann {et~al.}(2015)Ackermann, Ajello, Albert, Atwood, Baldini,
  Ballet, Barbiellini, Bastieri, Gonzalez, Bellazzini, Bissaldi, Blandford,
  Bloom, Bonino, Bottacini, Bregeon, Bruel, Buehler, Buson, Caliandro, Cameron,
  Caputo, Caragiulo, Caraveo, Cavazzuti, Cecchi, Chekhtman, Chiang, Chiaro,
  Ciprini, Cohen-Tanugi, Conrad, Cutini, D’Ammando, de~Angelis, de~Palma,
  Desiante, Venere, Domi´nguez, Drell, Favuzzi, Fegan, Ferrara, Focke,
  Fuhrmann, Fukazawa, Fusco, Gargano, Gasparrini, Giglietto, Giommi, Giordano,
  Giroletti, Godfrey, Green, Grenier, Grove, Guiriec, Harding, Hays, Hewitt,
  Hill, Horan, Jogler, Jóhannesson, Johnson, Kamae, Kuss, Larsson, Latronico,
  Li, Li, Longo, Loparco, Lott, Lovellette, Lubrano, Magill, Maldera, Manfreda,
  Max-Moerbeck, Mayer, Mazziotta, McEnery, Michelson, Mizuno, Monzani,
  Morselli, Moskalenko, Murgia, Nuss, Ohno, Ohsugi, Ojha, Omodei, Orlando,
  Ormes, Paneque, Pearson, Perkins, Perri, Pesce-Rollins, Petrosian, Piron,
  Pivato, Porter, Rainò, Rando, Razzano, Readhead, Reimer, Reimer, Schulz,
  Sgrò, Siskind, Spada, Spandre, Spinelli, Suson, Takahashi, Thayer, Thompson,
  Tibaldo, Torres, Tosti, Troja, Uchiyama, Vianello, Wood, Wood, Zimmer,
  Berdyugin, Corbet, Hovatta, Lindfors, Nilsson, Reinthal, Sillanpää,
  Stamerra, Takalo, \& Valtonen}]{Ackermann_2015}
Ackermann, M., Ajello, M., Albert, A., {et~al.} 2015, The Astrophysical Journal
  Letters, 813, L41

\bibitem[{Aharonian(2000)}]{Aharonian_2000}
Aharonian, F. 2000, New Astronomy, 5, 377

\bibitem[{{Aharonian, F.} {et~al.}(2008){Aharonian, F.}, {Akhperjanian, A. G.},
  {Barres de Almeida, U.}, {Bazer-Bachi, A. R.}, {Behera, B.}, {Beilicke, M.},
  {Benbow, W.}, {Bernlöhr, K.}, {Boisson, C.}, {Bolz, O.}, {Borrel, V.},
  {Braun, I.}, {Brion, E.}, {Brown, A. M.}, {Bühler, R.}, {Bulik, T.},
  {Büsching, I.}, {Boutelier, T.}, {Carrigan, S.}, {Chadwick, P. M.},
  {Chounet, L.-M.}, {Clapson, A. C.}, {Coignet, G.}, {Cornils, R.},
  {Costamante, L.}, {Dalton, M.}, {Degrange, B.}, {Dickinson, H. J.},
  {Djannati-Ataï, A.}, {Domainko, W.}, {Drury, L. O'C.}, {Dubois, F.}, {Dubus,
  G.}, {Dyks, J.}, {Egberts, K.}, {Emmanoulopoulos, D.}, {Espigat, P.},
  {Farnier, C.}, {Feinstein, F.}, {Fiasson, A.}, {Förster, A.}, {Fontaine,
  G.}, {Funk, Seb.}, {Füßling, M.}, {Gallant, Y. A.}, {Giebels, B.},
  {Glicenstein, J. F.}, {Glück, B.}, {Goret, P.}, {Hadjichristidis, C.},
  {Hauser, D.}, {Hauser, M.}, {Heinzelmann, G.}, {Henri, G.}, {Hermann, G.},
  {Hinton, J. A.}, {Hoffmann, A.}, {Hofmann, W.}, {Holleran, M.}, {Hoppe, S.},
  {Horns, D.}, {Jacholkowska, A.}, {de Jager, O. C.}, {Jung, I.},
  {Katarzyński, K.}, {Kendziorra, E.}, {Kerschhaggl, M.}, {Khélifi, B.},
  {Keogh, D.}, {Komin, Nu.}, {Kosack, K.}, {Lamanna, G.}, {Latham, I. J.},
  {Lemière, A.}, {Lemoine-Goumard, M.}, {Lenain, J.-P.}, {Lohse, T.}, {Martin,
  J. M.}, {Martineau-Huynh, O.}, {Marcowith, A.}, {Masterson, C.}, {Maurin,
  D.}, {Maurin, G.}, {McComb, T. J. L.}, {Moderski, R.}, {Moulin, E.}, {de
  Naurois, M.}, {Nedbal, D.}, {Nolan, S. J.}, {Ohm, S.}, {Olive, J.-P.}, {de
  Oña Wilhelmi, E.}, {Orford, K. J.}, {Osborne, J. L.}, {Ostrowski, M.},
  {Panter, M.}, {Pedaletti, G.}, {Pelletier, G.}, {Petrucci, P.-O.}, {Pita,
  S.}, {Pühlhofer, G.}, {Punch, M.}, {Ranchon, S.}, {Raubenheimer, B. C.},
  {Raue, M.}, {Rayner, S. M.}, {Renaud, M.}, {Ripken, J.}, {Rob, L.}, {Rolland,
  L.}, {Rosier-Lees, S.}, {Rowell, G.}, {Rudak, B.}, {Ruppel, J.}, {Sahakian,
  V.}, {Santangelo, A.}, {Schlickeiser, R.}, {Schöck, F.}, {Schröder, R.},
  {Schwanke, U.}, {Schwarzburg, S.}, {Schwemmer, S.}, {Shalchi, A.}, {Sol, H.},
  {Spangler, D.}, {Stawarz, Ł.}, {Steenkamp, R.}, {Stegmann, C.}, {Superina,
  G.}, {Tam, P. H.}, {Tavernet, J.-P.}, {Terrier, R.}, {van Eldik, C.},
  {Vasileiadis, G.}, {Venter, C.}, {Vialle, J. P.}, {Vincent, P.}, {Vivier,
  M.}, {Völk, H. J.}, {Volpe, F.}, {Wagner, S. J.}, {Ward, M.}, {Zdziarski, A.
  A.}, \& {Zech, A.}}]{Aharonian_2008}
{Aharonian, F.}, {Akhperjanian, A. G.}, {Barres de Almeida, U.}, {et~al.} 2008,
  A\&A, 477, 481

\bibitem[{{Aharonian, F.} {et~al.}(2006){Aharonian, F.}, {Akhperjanian, A. G.},
  {Bazer-Bachi, A. R.}, {Beilicke, M.}, {Benbow, W.}, {Berge, D.}, {Bernlöhr,
  K.}, {Boisson, C.}, {Bolz, O.}, {Borrel, V.}, {Braun, I.}, {Breitling, F.},
  {Brown, A. M.}, {Bühler, R.}, {Carrigan, S.}, {Chadwick, P. M.}, {Chounet,
  L.-M.}, {Cornils, R.}, {Costamante, L.}, {Degrange, B.}, {Dickinson, H. J.},
  {Djannati-Ataï, A.}, {Drury, L. O'C.}, {Dubus, G.}, {Egberts, K.},
  {Emmanoulopoulos, D.}, {Espigat, P.}, {Feinstein, F.}, {Fontaine, G.}, {Funk,
  S.}, {Gallant, Y. A.}, {Giebels, B.}, {Glicenstein, J. F.}, {Goret, P.},
  {Hadjichristidis, C.}, {Hauser, D.}, {Hauser, M.}, {Heinzelmann, G.}, {Henri,
  G.}, {Hermann, G.}, {Hinton, J. A.}, {Hofmann, W.}, {Holleran, M.}, {Horns,
  D.}, {Jacholkowska, A.}, {de Jager, O. C.}, {Khélifi, B.}, {Komin, Nu.},
  {Konopelko, A.}, {Latham, I. J.}, {Le Gallou, R.}, {Lemière, A.},
  {Lemoine-Goumard, M.}, {Lohse, T.}, {Martin, J. M.}, {Martineau-Huynh, O.},
  {Marcowith, A.}, {Masterson, C.}, {McComb, T. J. L.}, {de Naurois, M.},
  {Nedbal, D.}, {Nolan, S. J.}, {Noutsos, A.}, {Orford, K. J.}, {Osborne, J.
  L.}, {Ouchrif, M.}, {Panter, M.}, {Pelletier, G.}, {Pita, S.}, {Pühlhofer,
  G.}, {Punch, M.}, {Raubenheimer, B. C.}, {Raue, M.}, {Rayner, S. M.},
  {Reimer, A.}, {Reimer, O.}, {Ripken, J.}, {Rob, L.}, {Rolland, L.}, {Rowell,
  G.}, {Sahakian, V.}, {Saugé, L.}, {Schlenker, S.}, {Schlickeiser, R.},
  {Schuster, C.}, {Schwanke, U.}, {Siewert, M.}, {Sol, H.}, {Spangler, D.},
  {Steenkamp, R.}, {Stegmann, C.}, {Superina, G.}, {Tavernet, J.-P.}, {Terrier,
  R.}, {Théoret, C. G.}, {Tluczykont, M.}, {van Eldik, C.}, {Vasileiadis, G.},
  {Venter, C.}, {Vincent, P.}, {Völk, H. J.}, {Wagner, S. J.}, \& {Ward,
  M.}}]{Aharonian_2006}
{Aharonian, F.}, {Akhperjanian, A. G.}, {Bazer-Bachi, A. R.}, {et~al.} 2006,
  A\&A, 448, L19

\bibitem[{Albert {et~al.}(2006)Albert, Aliu, Anderhub, Antoranz, Armada,
  Baixeras, Barrio, Bartko, Bastieri, Becker, Bednarek, Berger, Bigongiari,
  Biland, Bock, Bordas, Bosch-Ramon, Bretz, Britvitch, Camara, Carmona,
  Chilingarian, Ciprini, Coarasa, Commichau, Contreras, Cortina, Curtef,
  Danielyan, Dazzi, Angelis, de~los Reyes, Lotto, Domingo-Santamaría, Dorner,
  Doro, Errando, Fagiolini, Ferenc, Fernández, Firpo, Flix, Fonseca, Font,
  Fuchs, Galante, Garczarczyk, Gaug, Giller, Goebel, Hakobyan, Hayashida,
  Hengstebeck, Höhne, Hose, Hsu, Jacon, Jogler, Kalekin, Kosyra, Kranich,
  Kritzer, Laille, Liebing, Lindfors, Lombardi, Longo, López, López, Lorenz,
  Majumdar, Maneva, Mannheim, Mansutti, Mariotti, Martínez, Mazin, Merck,
  Meucci, Meyer, Miranda, Mirzoyan, Mizobuchi, Moralejo, Nilsson, Ninkovic,
  Oña-Wilhelmi, Otte, Oya, Paneque, Paoletti, Paredes, Pasanen, Pascoli,
  Pauss, Pegna, Persic, Peruzzo, Piccioli, Poller, Puchades, Prandini, Raymers,
  Rhode, Ribó, Rico, Rissi, Robert, Rügamer, Saggion, Sánchez, Sartori,
  Scalzotto, Scapin, Schmitt, Schweizer, Shayduk, Shinozaki, Sidro,
  Sillanpää, Sobczynska, Stamerra, Stark, Takalo, Temnikov, Tescaro, Teshima,
  Tonello, Torres, Turini, Vankov, Vitale, Wagner, Wibig, Wittek, Zanin, \&
  Zapatero}]{Albert_2007}
Albert, J., Aliu, E., Anderhub, H., {et~al.} 2006, The Astrophysical Journal,
  654, L119

\bibitem[{Aleksić {et~al.}(2015)Aleksić, Ansoldi, Antonelli, Antoranz, Babic,
  Bangale, Barrio, Becerra~González, Bednarek, Bernardini, Biasuzzi, Biland,
  Blanch, Bonnefoy, Bonnoli, Borracci, Bretz, Carmona, Carosi, Colin, Colombo,
  Contreras, Cortina, Covino, Da~Vela, Dazzi, De~Angelis, De~Caneva, De~Lotto,
  de~Oña~Wilhelmi, Delgado~Mendez, Dominis~Prester, Dorner, Doro, Einecke,
  Eisenacher, Elsaesser, Fidalgo, Fonseca, Font, Frantzen, Fruck, Galindo,
  García~López, Garczarczyk, Garrido~Terrats, Gaug, Godinović,
  González~Muñoz, Gozzini, Hadasch, Hanabata, Hayashida, Herrera, Hildebrand,
  Hose, Hrupec, Idec, Kadenius, Kellermann, Knoetig, Kodani, Konno, Krause,
  Kubo, Kushida, La~Barbera, Lelas, Lewandowska, Lindfors, Lombardi, Longo,
  López, López-Coto, López-Oramas, Lorenz, Lozano, Makariev, Mallot, Maneva,
  Mannheim, Maraschi, Marcote, Mariotti, Martínez, Mazin, Menzel, Miranda,
  Mirzoyan, Moralejo, Munar-Adrover, Nakajima, Neustroev, Niedzwiecki, Nilsson,
  Nishijima, Noda, Orito, Overkemping, Paiano, Palatiello, Paneque, Paoletti,
  Paredes, Paredes-Fortuny, Persic, Poutanen, Prada~Moroni, Prandini, Puljak,
  Reinthal, Rhode, Ribó, Rico, Rodriguez~Garcia, Rügamer, Saito, Saito,
  Satalecka, Scalzotto, Scapin, Schultz, Schweizer, Sillanpää, Sitarek,
  Snidaric, Sobczynska, Spanier, Stamerra, Steinbring, Storz, Strzys, Takalo,
  Takami, Tavecchio, Temnikov, Terzić, Tescaro, Teshima, Thaele, Tibolla,
  Torres, Toyama, Treves, Vogler, Will, Zanin, Collaboration), D'Ammando,
  Buson, (for~the Fermi-LAT~Collaboration), Lähteenmäki, Tornikoski, Hovatta,
  Readhead, Max-Moerbeck, \& Richards}]{Aleksic_2015}
Aleksić, J., Ansoldi, S., Antonelli, L.~A., {et~al.} 2015, Monthly Notices of
  the Royal Astronomical Society, 450, 4399

\bibitem[{Dhiman {et~al.}(2021)Dhiman, Gupta, Gaur, \& Wiita}]{Dhiman_2021}
Dhiman, V., Gupta, A.~C., Gaur, H., \& Wiita, P.~J. 2021, Monthly Notices of
  the Royal Astronomical Society, 506, 1198

\bibitem[{Domínguez {et~al.}(2011)Domínguez, Primack, Rosario, Prada,
  Gilmore, Faber, Koo, Somerville, Pérez-Torres, Pérez-González, Huang,
  Davis, Guhathakurta, Barmby, Conselice, Lozano, Newman, \&
  Cooper}]{Dominguez_2011}
Domínguez, A., Primack, J.~R., Rosario, D.~J., {et~al.} 2011, Monthly Notices
  of the Royal Astronomical Society, 410, 2556

\bibitem[{{Donath} {et~al.}(2023){Donath}, {Terrier}, {Remy}, {Sinha}, {Nigro},
  {Pintore}, {Kh\'elifi}, {Olivera-Nieto}, {Ruiz}, {Br\"ugge}, {Linhoff},
  {Contreras}, {Acero}, {Aguasca-Cabot}, {Berge}, {Bhattacharjee}, {Buchner},
  {Boisson}, {Carreto Fidalgo}, {Chen}, {de Bony de Lavergne}, {de Miranda
  Cardoso}, {Deil}, {F\"u\ss{}ling}, {Funk}, {Giunti}, {Hinton}, {Jouvin},
  {King}, {Lefaucheur}, {Lemoine-Goumard}, {Lenain}, {L\'opez-Coto},
  {Mohrmann}, {Morcuende}, {Panny}, {Regeard}, {Saha}, {Siejkowski},
  {Siemiginowska}, {Sip"ocz}, {Unbehaun}, {van Eldik}, {Vuillaume}, \&
  {Zanin}}]{gammapy:2023}
{Donath}, A., {Terrier}, R., {Remy}, Q., {et~al.} 2023, A\&A, 678, A157

\bibitem[{{Dorigo Jones} {et~al.}(2022){Dorigo Jones}, {Johnson}, {Muzahid},
  {Charlton}, {Chen}, {Narayanan}, {Sameer}, {Schaye}, \&
  {Wijers}}]{Jones_2022}
{Dorigo Jones}, J., {Johnson}, S.~D., {Muzahid}, S., {et~al.} 2022, \mnras,
  509, 4330

\bibitem[{Edelson {et~al.}(2002)Edelson, Turner, Pounds, Vaughan, Markowitz,
  Marshall, Dobbie, \& Warwick}]{Edelson_2002}
Edelson, R., Turner, T.~J., Pounds, K., {et~al.} 2002, The Astrophysical
  Journal, 568, 610

\bibitem[{{Foschini, L.} {et~al.}(2011){Foschini, L.}, {Ghisellini, G.},
  {Tavecchio, F.}, {Bonnoli, G.}, \& {Stamerra, A.}}]{Foschini_2011}
{Foschini, L.}, {Ghisellini, G.}, {Tavecchio, F.}, {Bonnoli, G.}, \& {Stamerra,
  A.} 2011, A\&A, 530, A77

\bibitem[{Giannios(2010)}]{Giannios_2010}
Giannios, D. 2010, Monthly Notices of the Royal Astronomical Society: Letters,
  408, L46

\bibitem[{{Green} {et~al.}(1986){Green}, {Schmidt}, \& {Liebert}}]{Green_1986}
{Green}, R.~F., {Schmidt}, M., \& {Liebert}, J. 1986, \apjs, 61, 305

\bibitem[{Liu {et~al.}(2023)Liu, Xue, Wang, Tan, \& Böttcher}]{Liu_2023}
Liu, R.-Y., Xue, R., Wang, Z.-R., Tan, H.-B., \& Böttcher, M. 2023, Monthly
  Notices of the Royal Astronomical Society, 526, 5054

\bibitem[{Marscher(2016)}]{Marscher_2016}
Marscher, A.~P. 2016, Galaxies, 4

\bibitem[{Marscher {et~al.}(2010)Marscher, Jorstad, Larionov, Aller, Aller,
  Lähteenmäki, Agudo, Smith, Gurwell, Hagen-Thorn, Konstantinova, Larionova,
  Larionova, Melnichuk, Blinov, Kopatskaya, Troitsky, Tornikoski, Hovatta,
  Schmidt, D’Arcangelo, Bhattarai, Taylor, Olmstead, Manne-Nicholas,
  Roca-Sogorb, Gómez, McHardy, Kurtanidze, Nikolashvili, Kimeridze, \&
  Sigua}]{Marscher_2010}
Marscher, A.~P., Jorstad, S.~G., Larionov, V.~M., {et~al.} 2010, The
  Astrophysical Journal Letters, 710, L126

\bibitem[{Petropoulou {et~al.}(2016)Petropoulou, Vasilopoulos, \&
  Giannios}]{Petropoulou_2016}
Petropoulou, M., Vasilopoulos, G., \& Giannios, D. 2016, Monthly Notices of the
  Royal Astronomical Society, 464, 2213

\bibitem[{{Raiteri} {et~al.}(2017){Raiteri}, {Villata}, {Acosta-Pulido},
  {Agudo}, {Arkharov}, {Bachev}, {Baida}, {Ben{\'\i}tez}, {Borman}, {Boschin},
  {Bozhilov}, {Butuzova}, {Calcidese}, {Carnerero}, {Carosati}, {Casadio},
  {Castro-Segura}, {Chen}, {Damljanovic}, {D'Ammando}, {di Paola},
  {Echevarr{\'\i}a}, {Efimova}, {Ehgamberdiev}, {Espinosa}, {Fuentes},
  {Giunta}, {G{\'o}mez}, {Grishina}, {Gurwell}, {Hiriart}, {Jermak}, {Jordan},
  {Jorstad}, {Joshi}, {Kopatskaya}, {Kuratov}, {Kurtanidze}, {Kurtanidze},
  {L{\"a}hteenm{\"a}ki}, {Larionov}, {Larionova}, {Larionova}, {L{\'a}zaro},
  {Lin}, {Malmrose}, {Marscher}, {Matsumoto}, {McBreen}, {Michel}, {Mihov},
  {Minev}, {Mirzaqulov}, {Mokrushina}, {Molina}, {Moody}, {Morozova},
  {Nazarov}, {Nikolashvili}, {Ohlert}, {Okhmat}, {Ovcharov}, {Pinna},
  {Polakis}, {Protasio}, {Pursimo}, {Redondo-Lorenzo}, {Rizzi},
  {Rodriguez-Coira}, {Sadakane}, {Sadun}, {Samal}, {Savchenko}, {Semkov},
  {Skiff}, {Slavcheva-Mihova}, {Smith}, {Steele}, {Strigachev}, {Tammi},
  {Thum}, {Tornikoski}, {Troitskaya}, {Troitsky}, {Vasilyev}, \&
  {Vince}}]{Raiteri_2017}
{Raiteri}, C.~M., {Villata}, M., {Acosta-Pulido}, J.~A., {et~al.} 2017, \nat,
  552, 374

\bibitem[{{Urry} \& {Padovani}(1995)}]{Urry_1995}
{Urry}, C.~M. \& {Padovani}, P. 1995, \pasp, 107, 803

\bibitem[{{Vaughan} {et~al.}(2003){Vaughan}, {Edelson}, {Warwick}, \&
  {Uttley}}]{Vaughan_2003}
{Vaughan}, S., {Edelson}, R., {Warwick}, R.~S., \& {Uttley}, P. 2003, \mnras,
  345, 1271

\end{thebibliography}
\bibliographystyle{aa}

\end{document}